# Nonreciprocal charge transport in an iron-based superconductor with broken inversion symmetry engineered by a hydrogen-concentration gradient

Takayuki Nagai,[1,*] Yukito Nishio,[1] Jumpei Matsumoto,[2] Kota Hanzawa,[2] Hidenori Hiramatsu,[2,3] Hideo Hosono,[3,4] Tsuyoshi Kimura[1]

[1]Quantum-Phase Electronics Center (QPEC) and Department of Applied Physics, University of Tokyo, Tokyo, Japan

[2]Materials and Structures Laboratory, Institute of Integrated Research, Institute of Science Tokyo, Yokohama, Japan

[3]MDX Research Center for Element Strategy, Institute of Integrated Research, Institute of Science Tokyo, Yokohama, Japan

[4]Research Center for Materials Nanoarchitectonics, National Institute for Materials Science, Tsukuba, Japan

*t-nagai@ap.t.u-tokyo.ac.jp

**The breaking of spatial inversion symmetry in condensed matter gives rise to intriguing physical properties, such as ferroelectricity, piezoelectricity, spin-momentum locking, and nonreciprocal responses. Here we propose that a concentration gradient, which often persists as a quasi-stable nonequilibrium state with long relaxation times in solids, can serve as a general platform for inversion symmetry breaking. This concept is demonstrated in an epitaxial thin film of the hydrogen-doped SmFeAsO (Sm1111:H) superconductor with a depthwise hydrogen-concentration gradient introduced by an optimized topotactic reaction. This film exhibits nonreciprocal charge transport—that is, current-direction-dependent resistance—which serves as a key signature of broken inversion symmetry. A large nonreciprocal signal emerges in the vicinity of the superconducting transition, which we attribute to vortex-motion nonreciprocity arising from an asymmetric pinning landscape created by the hydrogen-concentration gradient. Owing to the high critical temperature of Sm1111:H, vortex-origin nonreciprocity is observed above 40 K, which represents the highest temperature ever reported among single bulk materials without an artificially hetero-layered structure. Our findings establish concentration-gradient engineering as a versatile and chemically general route to creating inversion-broken states in otherwise centrosymmetric hosts, paving the way for a broader landscape of odd-parity-driven functionalities.**

**Introduction**

Inversion symmetry ($\mathcal{P}$) breaking is a fundamental ingredient in condensed matter, enabling practical functionalities and seeding exotic quantum phenomena. Representative examples include ferroelectricity and piezoelectricity, which arise from broken $\mathcal{P}$ and underpin the operating principles of non-volatile memories and sensors, and have driven widespread technological development [1, 2]. Concomitantly, $\mathcal{P}$-breaking activates antisymmetric spin-orbit coupling, which in turn permits spin-split band structures without magnetism, parity-mixed non-trivial superconductivity, and nonreciprocal transport and optical properties [3, 4, 5, 6, 7, 8]. Realizing a $\mathcal{P}$-broken platform is therefore a central objective in materials science.

Traditionally, inversion asymmetry has been achieved via various routes, including intrinsically non-centrosymmetric crystal structures, surfaces/interfaces/heterostructures[9, 10, 11, 12], externally applied electric fields [13], and strain gradients [14, 15]. Each route faces practical constraints. Non-centrosymmetric compounds are comparatively scarcer than centrosymmetric ones, and $\mathcal{P}$-related domains often cancel their intrinsic responses unless a poling field is applied or electrodes are fabricated within single domains. Interface-based approaches are limited by material combinations and are sensitive to interfacial morphology. Field-induced polarity frequently demands complex device structures (e.g., electric double-layer transistors), while strain-gradient-induced displacements are typically small. These limitations motivate a more universal and controllable platform for $\mathcal{P}$-breaking.

Herein, we propose a symmetry-based strategy in which a concentration gradient of dopant ions plays a crucial role in breaking $\mathcal{P}$. In condensed matter, a concentration gradient frequently arises as a quasi-stable state with long relaxation times. By symmetry, a concentration gradient transforms as a polar vector identically to an electric polarization, and therefore breaks $\mathcal{P}$. Once such a gradient is established, phenomena characteristic of $\mathcal{P}$-breaking are expected to emerge. To demonstrate this concept, we engineered a depthwise concentration gradient of doped hydrogen in the iron-based superconductor

SmFeAsO$_{1-x}$H$_x$ (Sm1111:H) and examined nonreciprocal charge transport, a benchmark electrical signature that necessitates broken $\mathcal{P}$. Engineering a concentration gradient thus converts a centrosymmetric host into a $\mathcal{P}$-broken platform. More broadly, this approach provides a general and chemically versatile paradigm for creating and controlling $\mathcal{P}$-breaking in condensed-matter systems.

Consider a spatial distribution $n(\mathbf{r})$ of dopant or constituent species with a finite gradient $\nabla n$. Treating the plane perpendicular to the gradient as uniform (neglecting in-plane rotational anisotropy), the characters (+1 for even, −1 for odd) of $\nabla n$ under the symmetry operations $\{2_\parallel, 2_\perp, m_\perp, m_\parallel\}$ are $\{+1, -1, -1, +1\}$, identical to those of the electric polarization $\boldsymbol{P}$ (Fig. 1a). In other words, a system endowed with a concentration gradient $\nabla n$ behaves as a polar system with $\boldsymbol{P} \parallel \nabla n$. Such gradients arise naturally and ubiquitously in solids during processes involving ionic diffusion, notably topotactic reactions, and are common rather than exceptional [16, 17, 18, 19, 20, 21, 22]. Although the resulting state is formally non-equilibrium, diffusion barriers can render its relaxation exceedingly slow; it often exists as a glass-like quasi-stable state with long lifetime on experimental timescales. Therefore, the polar state is long-lived. Crucially, this approach is independent of the parent crystal symmetry: even centrosymmetric hosts can be rendered polar via $\nabla n$. Moreover, when $\nabla n$ is produced topotactically, the magnitude of $\nabla n$ (hence the degree of $\mathcal{P}$-breaking) can be tuned continuously by electrochemical and thermal annealing.

To demonstrate this concept, we chose SmFeAsO (Sm1111) as a test platform. As shown in Fig. 1b, Sm1111 comprises alternating SmO and FeAs layers in a centrosymmetric crystal structure (space group: *Cmma* at temperatures below 130 K) [23]. The undoped parent phase is a metallic antiferromagnet, while electron doping by substituting F$^-$ or H$^-$ for the O$^{2-}$ site induces superconductivity [24, 25]. Sm1111 exhibits the highest superconducting transition temperature among the 1111 family of iron-based superconductors ($T_c \approx 55$ K for bulk [26]). Notably, epitaxial films can be electron-doped by topotactic hydrogen substitution: Sm1111 epitaxially grown on an MgO substrate and annealed in CaH$_2$ powder undergoes $O^{2-} \leftrightarrow H^-$ exchange, enabling wide control of the H content (from a few to several tens of percent) and $T_c$ (up to ~ 48 K) [27]. Under appropriate conditions, a through-thickness $H^-$ concentration gradient emerges naturally. In such films, the host crystal lattice remains centrosymmetric, yet $\nabla n_H$ defines a polar axis and breaks $\mathcal{P}$, making Sm1111:H an ideal testbed for our strategy.

To detect $\mathcal{P}$-breaking, we employ nonreciprocal charge transport (NCT), a directional dichroism in charge transport that strictly requires broken $\mathcal{P}$ [28, 29, 8]. In polar systems, the electrical voltage drop is phenomenologically described as a function of magnetic field $B$ and the current $I$ by

$$V = R_0 I (1 + \beta B^2 + \gamma (\boldsymbol{B} \times \hat{\boldsymbol{z}}) \cdot \boldsymbol{I}) \tag{1}$$

where $\hat{\boldsymbol{z}}$ denotes the unit vector along the polar axis [29, 30, 31, 32]. The second term represents ordinary magnetoresistance. The third term depends on the current direction and encodes nonreciprocal transport. If $\mathcal{P}$ is preserved, the symmetry forbids this term; therefore, observing nonreciprocal transport serves as an electrical signature of $\mathcal{P}$-breaking. Practically, since the third term appears in measured voltage as a nonlinear response proportional to $I^2$, the nonreciprocal coefficient $\gamma$ is clearly extracted by using second-harmonic resistance $R^{2\omega} = \gamma R_0 BI/2$ under ac current and in-plane magnetic fields, most sensitively for $\boldsymbol{I} \perp \boldsymbol{B}$ (see Methods). Moreover, in superconductors, the nonreciprocity is strongly amplified near the superconducting transition via superconducting fluctuations (paraconductivity) [33, 34, 35] and vortex dynamics [34, 36, 37, 38, 39]. The Sm1111 epitaxial thin film with the hydrogen gradient $\nabla n_H$ is therefore ideally suited to demonstrate the $\mathcal{P}$-breaking induced by the concentration gradient by detecting nonreciprocal charge transport (Fig. 1c).

**Results**

Sm1111:H epitaxial thin films were prepared following procedures established in previous reports [40, 27]. The *c*-axis-oriented undoped Sm1111 films were first grown on MgO(001) substrates by pulsed laser deposition (PLD), after which oxygen was partially replaced by hydrogen via topotactic post-annealing in CaH$_2$ powder. In this study, we fabricated two films with distinct hydrogen distributions: sample #1 (thickness 60 nm) and sample #2 (68 nm). Figure 1d shows the depth profiles of hydrogen concentration *x* measured by secondary-ion mass spectrometry (SIMS). In sample #2, the hydrogen level is nearly uniform through the thickness with an average $x \approx 0.04$. In contrast, sample #1 exhibits a pronounced through-thickness gradient: $x \approx 0.14$ near the top surface, decreasing to $x \approx 0.04$ close to the Sm1111/MgO interface. Note that the apparent spike at the surface and the perturbation near the interface, where the composition changes abruptly, are not intrinsic; they are attributable to a well-known SIMS pile-up effect [41]. Additionally, the vicinity of the MgO substrate is susceptible to charging due to its highly insulating nature. Although the H doping up to ~35% and a maximum $T_c$ of ~48 K are achievable under optimised annealing conditions [27], the present study deliberately employed lightly doped films to facilitate unambiguous verification of the H concentration gradient via transport measurements. Because the $T_c$-*x* phase diagram of Sm1111:H is dome-shaped [25, 42], $T_c$ varies little with *x* near the dome centre; thus, the H gradient has little apparent effect on the behaviour of the superconducting transition there. In contrast, on the underdoped side (the dome edge), the $T_c$ depends sensitively on *x*, so the H concentration gradient is expected to manifest as a multi-step superconducting transition. Figure 1e displays the temperature dependence of the electrical resistivity *ρ*(*T*) for sample #1. A clear metallic behaviour is observed in the normal state, and a drop of *ρ* indicating superconducting transition appears with $T_c^{\text{onset}} = 41$ K. Although applying a magnetic field perpendicular to the *c* axis shifts the $T_c$ slightly to lower temperatures, the superconductivity persists up to 9 T, reflecting an extremely high upper critical field and reproducing earlier reports [27, 43, 44]. The inset of Fig. 1e compares *ρ* (*T*) for samples #1 and #2 on a logarithmic scale. At first glance, the two films appear to display similar *ρ*-*T* curves with nearly identical $T_c$ values (see

Supplementary Note 4). The uniform film (#2) shows a sharp fall near $T_c^{\text{onset}} = 41.5$ K, whereas the gradient film (#1) exhibits a multi-step decrease with shoulder-like structures around the transition, which is consistent with a distribution of local $T_c$ values arising from the hydrogen gradient. As summarized in Fig. 1f, the $T_c$ values extracted from the $\rho$-$T$ curves are plotted on the established phase diagram. While the $T_c$ values are slightly reduced compared with bulk, the $T_c^{\text{onset}}$ values of samples #1 and #2 track the superconducting dome reasonably well.

We next focus on the charge transport properties under ac excitation. Gold electrodes for conventional four-terminal measurements were deposited by sputtering on the prepared films, and Au wires were bound using silver paste; no special nanofabrication such as FIB processing or electrode patterning was employed. Unless otherwise noted, we applied an ac current of $I_{\text{ac}} = 0.5$ mA (current density $j_{\text{ac}} \approx 10^2$ A/cm$^2$) and measured the first- and second-harmonic resistances, $R_{xx}^{\omega}$ and $R_{xx}^{2\omega}$, with a lock-in amplifier. Figures 2a and 2b show the magnetic-field dependence of $R_{xx}^{\omega}$ and $R_{xx}^{2\omega}$, respectively, for sample #1 at various temperatures across $T_c$. The magnetic field $\boldsymbol{B}$ and current $\boldsymbol{I}$ were applied mutually orthogonally and both perpendicular to the hydrogen concentration gradient, i.e., $\boldsymbol{B} \perp \boldsymbol{I} \perp \boldsymbol{\nabla} n_{\text{H}}$ (see inset schematic). In the normal state well above $T_c$, changes in both $R_{xx}^{\omega}$ and $R_{xx}^{2\omega}$ were too small to be detected by our experimental setup. Approaching the transient regime, $R_{xx}^{\omega}$ displays a large positive magnetoresistance, reflecting the $B$-induced suppression of superconductivity. At the same temperature, $R_{xx}^{2\omega}$ becomes finite and exhibits an antisymmetric peak-valley structure centered at $B = 0$ T. This line shape is the canonical fingerprint of nonreciprocal charge transport in superconductors, as reported previously [33, 34, 45]. The observation of the finite $R_{xx}^{2\omega}$ thus indicates that sample #1 is characterized by broken $\mathcal{P}$. Figure 2c presents $R_{xx}^{2\omega}$-$B$ curves at 33 K, where the signal is the most pronounced, for various current amplitudes. The antisymmetric structure grows monotonically with increasing $I$. The current-amplitude dependence of $R_{xx}^{2\omega}$ at the peak-valley field ($B = 1.95$ T) is summarized in Fig. 2d: in the low-current regime ($\lesssim 0.5$ mA; inset) $R_{xx}^{2\omega}$ increases linearly with $I_{ac}$, while at higher currents it deviates from the linear relationship, plausibly due to Joule heating, current-induced suppression of superconductivity, and/or higher-order effect. These results demonstrate that $R_{xx}^{2\omega} \propto BI$, i.e., the expected scaling for nonreciprocal transport, holds in the low-magnetic-field and low-current limit. Furthermore, the observation that $R_{xx}^{2\omega}$ appears exclusively in the vicinity of the superconducting transition and remains absent in the normal state clearly rules out thermoelectric artifacts such as the Nernst effect caused by Joule heating at a current contact [46, 47]. This finding indicates that the nonreciprocal charge transport observed in sample #1 is intimately associated with the superconducting transition.

To identify the origin of $\mathcal{P}$-breaking, we measured $R_{xx}^{\omega}$ and $R_{xx}^{2\omega}$ under several current and magnetic field geometries. Figures 3a, b show the magnetic field dependence of $R_{xx}^{\omega}$ and $R_{xx}^{2\omega}$, respectively, in three configurations: (i) $\boldsymbol{I} \perp \boldsymbol{B} \perp \boldsymbol{\nabla} n_H$, (ii) $\boldsymbol{I} \perp \boldsymbol{B} \parallel \boldsymbol{\nabla} n_H$, and (iii) $\boldsymbol{I} \parallel \boldsymbol{B} \perp \boldsymbol{\nabla} n_H$. Although the amplitudes differ, $R_{xx}^{\omega}$ shows a positive magnetoresistance in all the three geometries. In contrast, a pronounced antisymmetric peak-valley structure in $R_{xx}^{2\omega}$ appeared only for geometry (i). Weak residual $R_{xx}^{2\omega}$ signals in the other geometries may be attributed to slight misalignment. This clear geometry selectivity accords with the selection rules of Eq. (1): with the polar axis $\boldsymbol{z} \parallel \boldsymbol{\nabla} n_H$, the nonreciprocal term $\gamma(\boldsymbol{B} \times \boldsymbol{z}) \cdot \boldsymbol{I}$ is maximised when $\boldsymbol{I} \parallel (\boldsymbol{B} \times \boldsymbol{z})$ (i.e., $\boldsymbol{I} \perp \boldsymbol{B} \perp \boldsymbol{z}$) and vanishes for $\boldsymbol{B} \parallel \boldsymbol{z}$ or $\boldsymbol{I} \parallel \boldsymbol{B}$. These results indicate that sample #1 realises a polar $\mathcal{P}$-broken state that is symmetry equivalent to an electric polarization $\boldsymbol{P}$ aligned with $\boldsymbol{\nabla} n_H$.

Figures 3c, d compare $R_{xx}^{\omega}$ and $R_{xx}^{2\omega}$ for the two films in a fixed geometry, where $\boldsymbol{B}$, $\boldsymbol{I}$ and the film normal (thickness direction) are mutually orthogonal (see inset figures). Both samples exhibit positive magnetoresistance in $R_{xx}^{\omega}$. However, a robust $R_{xx}^{2\omega}$ signal is found only in sample #1 (with a hydrogen concentration gradient), while no $R_{xx}^{2\omega}$ is detected in sample #2 (without a gradient). This contrast demonstrates that the nonreciprocal charge transport originates from $\boldsymbol{\nabla} n_H$. Strictly speaking, the surface

and Sm1111/MgO interface also break inversion symmetry locally, but the null $R_{xx}^{2\omega}$ result in sample #2 excludes them as dominant sources. The observed nonreciprocity is therefore governed primarily by the hydrogen concentration gradient.

**Discussion**

To address the mechanism of the nonreciprocal charge transport in sample #1, we qualitatively evaluate the nonreciprocal coefficient $\gamma$. Several definitions exist in the literature [34, 36, 48]; here we adopt the form derived from Eq. (1): $\gamma = \frac{2R_{xx}^{2\omega}}{R_{xx}^{\omega}BI_0}$, which is evaluated from the slope of $R_{xx}^{2\omega}/R_{xx}^{\omega}$-$B$ curve around 0 T following previous studies [36, 45] (see Supplementary Note 2). In superconductors, there are two principal mechanisms for nonreciprocal charge transport especially formulated for two-dimensional (2D) superconductors: (I) nonreciprocal paraconductivity governed by the amplitude fluctuations of the superconducting order parameter [33, 35, 8], and (II) vortex-motion-driven nonreciprocity [34, 36, 37, 38, 39, 48]. In scenario (I), $\gamma$ increases upon approaching the mean-field transition temperature $T_{c0}$ and is maximal on the higher-temperature side of the transient regime. In scenario (II), $\gamma$ grows toward the Berezinskii‑Kosterlitz‑Thouless (BKT) temperature $T_{BKT}$, where vortices and antivortices bind into pairs, and is maximal to the lower-temperature side of the transient regime. Our Sm1111:H films are bulk superconductors with thickness $d \approx 50$ nm. Reported coherence lengths $\xi_c$ of Sm1111:H epitaxial thin films are a few nanometers [27]. Therefore, $d \gg \xi_c$ and the order parameter is effectively three-dimensional (3D). Consequently, order parameter fluctuations are suppressed, and a dominant contribution from paraconductivity is unlikely. Although the BKT transition is absent due to the 3D nature, an applied magnetic field readily creates vortex lines, and vortex motion can provide an active route to nonreciprocity. Figures 4a,b summarise the temperature dependences of $\rho_{xx}^{\omega}$ and $\gamma'$, which is the size-independent coefficient defined as $\gamma/S$ ($S$: cross sectional area). The $\gamma'$ begins to develop at $T_c^{onset}$ and then grows steeply toward lower temperatures within the transient regime. This trend favours vortex motion over paraconductivity as the primary origin of the nonreciprocal response. A microscopic picture is as follows. An in-plane magnetic field nucleates vortex lines (for example, the lower critical field $H_{c1}$ in polycrystalline SmFeAsO$_{1-x}$F$_x$ is ~13 mT [49], and in-plane $H_{c1}$ of single crystalline SmFe$_{0.92}$Co$_{0.08}$AsO is ~4 mT [50], indicating that vortices are easily generated.). When an in-plane current perpendicular to the field is applied, the Lorentz force drives vortex motion along the film normal. The hydrogen concentration gradient produces an asymmetric potential along $\nabla n_H$, so the vortex mobility differs for forward versus backward motion along this axis. This mobility asymmetry yields a finite second-harmonic voltage and manifests as nonreciprocity, i.e. $R(I, B) \neq R(-I, B)$ and/or $R(I, B) \neq R(I, -B)$. The nonreciprocal charge transport arising from the imbalance of vortex motion has been observed in other asymmetric systems [38, 51].

Figure 4c compares the temperature evolution of $\gamma'$ in gradient-engineered Sm1111 with superconductors showing nonreciprocal charge transport arising from vortex motion. In prior materials, the relatively low $T_c$ (typically < 10 K) confines detectable $\gamma'$ to cryogenic temperatures, and $\gamma'$ generally grows upon cooling. In contrast, the high-$T_c$ Sm1111 with $\nabla n_H$ exhibits nonreciprocity already from ~40 K, with comparatively large $\gamma'$, markedly deviating from the established trend. Notably, Sm1111:H shows the highest onset temperature of nonreciprocity reported to date among single-bulk materials lacking artificial heterostructures [52]. These findings indicate that concentration-gradient engineering offers a practical route to $\mathcal{P}$-breaking that produces robust nonreciprocal responses at elevated temperatures.

**Conclusion**

In conclusion, we have proposed a symmetry-based strategy in which a concentration gradient $\nabla n$ can serve as a new platform for realising the inversion symmetry ($\mathcal{P}$)-broken state. We have demonstrated this concept by observing the benchmark signature of $\mathcal{P}$-breaking—nonreciprocal charge transport—in SmFeAsO (Sm1111) epitaxial thin films with a hydrogen concentration gradient $\nabla n_\mathrm{H}$. From the geometry selectivity of the second-harmonic response and the $\gamma$-$T$ curve, we conclude that the dominant origin of the nonreciprocal charge transport in Sm1111 with $\nabla n_\mathrm{H}$ is vortex-motion nonreciprocity arising from an asymmetric pinning potential created by $\nabla n_\mathrm{H}$. We further note that, unlike conventional Sm1111, $\mathcal{P}$-breaking introduced by $\nabla n_\mathrm{H}$ might allow a unique superconducting state typified by the singlet-triplet mixing, as discussed in non-centrosymmetric superconductors [6, 53].

In the present study, the concentration gradient was realised via a topotactic reaction that introduced an asymmetric dopant profile. Since topotactic ion-exchange routes including isotope substitution [54] and electrochemical insertion/extraction [55] are well established in solid-state science, our approach is broadly applicable to a wide range of material systems. While our focus was on nonreciprocal transport in superconductors, the same symmetry principle should enable optical second-harmonic generation, piezoelectric responses, and other inversion-symmetry-odd phenomena in gradient-engineered platforms. More broadly, concentration-gradient engineering offers a versatile and chemically general route to creating $\mathcal{P}$-broken states−opening a pathway to a wider landscape of functional materials.

**Methods**

Sample preparation

The undoped Sm1111 heteroepitaxial thin films were grown on (001)-oriented MgO single-crystalline substrates (size: 10 mm × 10 mm × 0.5 mm) by PLD using the second harmonic ($\lambda$ = 532 nm) of an Nd:YAG laser. The Nd:YAG PLD-film-growth procedure followed the details reported in Refs. [40]. Hydrogen substitution in Sm1111 was achieved by topotactic ion exchange between as-grown epitaxial

Sm1111 films and CaH$_2$ powder. Further procedural details are provided in Ref. [27]. Depth profiles of the hydrogen concentration were obtained by SIMS at Toray Research Center, Inc.

Transport measurements

After the annealing procedure, the films were diced into bar-shaped specimens. Au electrodes for four-probe measurements were deposited, and Au wires were attached with conductive Ag paste using standard procedures. Notably, no micro-/nanofabrication such as FIB processing was employed for electrode patterning or wiring. The first- and second-harmonic resistances, $R^\omega$ and $R^{2\omega}$, were measured using an ac current source (Model 6221, Keithley) and lock-in amplifiers (LI5640, NF Corporation). Unless otherwise noted, the amplitude and frequency of the injected current were set to 0.5 mA (root mean square) and 13 Hz, respectively. All transport measurements were performed using a Physical Property Measurement System (PPMS, Quantum Design). During the ac resistance measurements, the phases of the first- (second-) harmonic signals were confirmed to be approximately 0 ($\pi/2$), consistent with the theoretical expectation (see Supplementary Note 1). All the $R^\omega$ (or $R^{2\omega}$) signals presented in the main text are $x$ (or $y$) components of the lock-in output, respectively. The $R^\omega$ and $R^{2\omega}$ data were symmetrised and antisymmetrised, respectively, with respect to the magnetic field $B$ (see Supplementary Note 4).

**Acknowledgements**

We thank H. Arisawa, N. D. Kahnh and T. Ideue for fruitful discussions and experimental supports. The present work was supported in part by KAKENHI (Grant Numbers JP24K08561, JP 25H01247, and JP 25H00392) and a research grant from the Murata Science and Education Foundation, (K.S.). H. Hiramatsu and H. Hosono were supported by the Ministry of Education, Culture, Sports, Science, and Technology (MEXT) through the Element Strategy Initiative to Form Core Research Center (grant no. JPMXP0112101001).

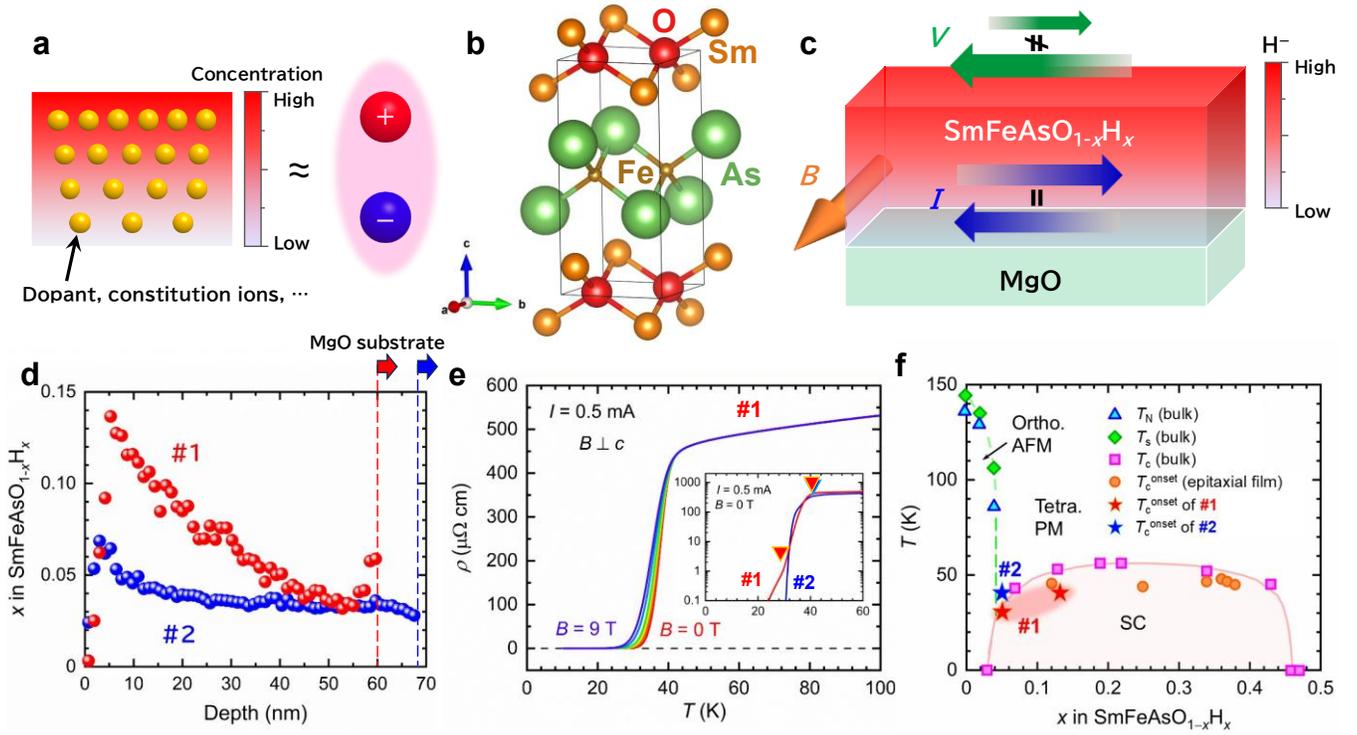

**Fig. 1 | Concentration-gradient-driven inversion-symmetry breaking and nonreciprocal charge transport, and basic properties of the prepared films. a**, Concept of inversion symmetry breaking driven by the concentration gradient $\nabla n$, which is symmetry-equivalent to an electric polarization $\boldsymbol{P}$. **b**, Schematic of the crystal structure of Sm1111; orange, red, brown, and green spheres indicate Sm, O, Fe, As, respectively. The image of the crystal structure was drawn using the software VESTA [56]. **c**, Schematic illustration of nonreciprocal charge transport in iron-based superconductor Sm1111 with hydrogen concentration gradient $\nabla n_{\mathrm{H^H}}$. $V$, $\boldsymbol{B}$, and $\boldsymbol{I}$ denote a voltage drop, a magnetic field, and an electric current, respectively. **d**, Depth profile of hydrogen content $x$ in $\mathrm{SmFeAsO_{1-x}H_x}$ measured by Second Ion Mass Spectrometry (SIMS). Dashed lines indicate the Sm1111/MgO interface. **e**, Temperature dependence of the electrical resistivity $\rho$ of sample #1 under in-plane magnetic field $B = 0, 1, 2, 3, 6,$ and 9 T. The applied current is 0.5 mA. The inset shows the logarithmic plot of $\rho$-$T$ curves of samples #1 and #2 under zero magnetic field. The filled triangles denote $T_c^{\mathrm{onset}}$, defined by the onset of the decrease in $\rho$.

In Sample #1, the triangle at lower temperatures represents the anomaly with shoulder-like structure in transient regime. **f**, Phase diagram of SmFeAsO$_{1-x}$H$_x$ as a function of $x$. Filled triangles, diamonds, and squares indicate the Néel temperature $T_N$, structural phase transition temperature $T_S$, and superconducting transition temperature $T_c$ of bulk samples, respectively, as reported in Refs. [25, 42]. Filled circles mark $T_c^{onset}$ for epitaxial thin films prepared by the same technique [27, 44]. Red and blue stars represent $T_c^{onset}$ of samples #1 and #2, respectively. Two stars of sample #1 mark the temperature at which a multi-step drop appears in the transient region of the $\rho$-$T$ curve.

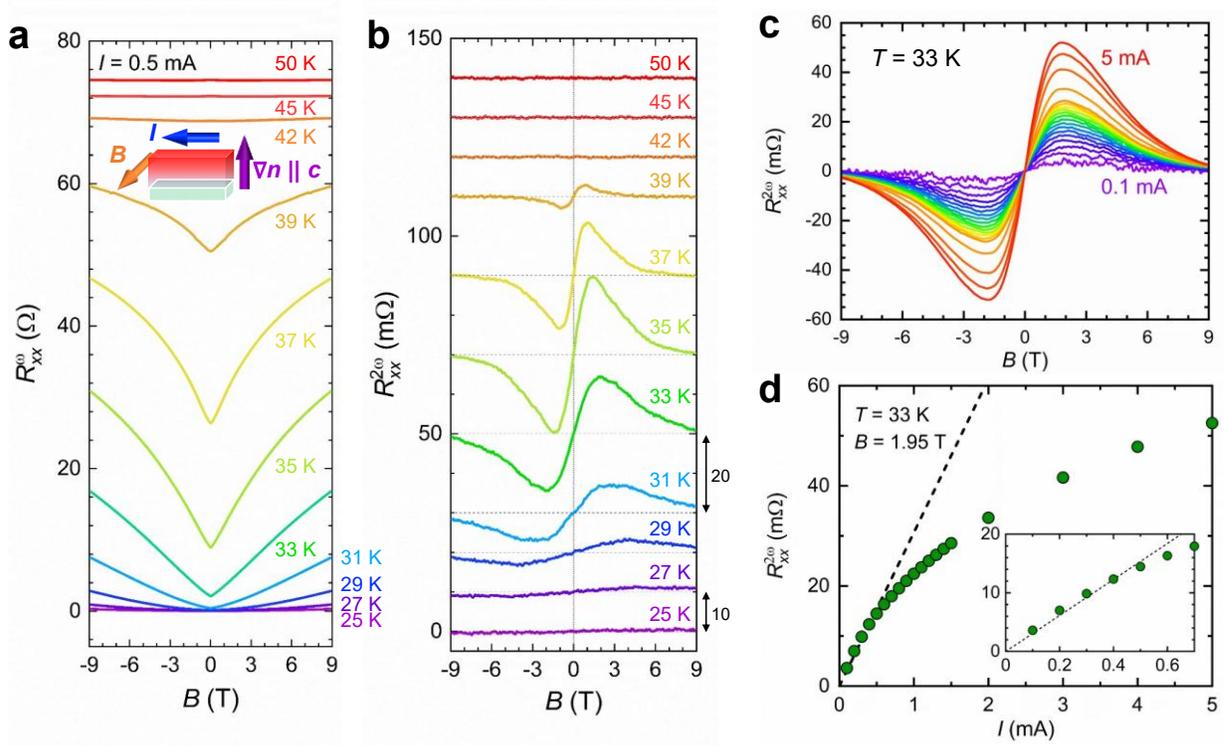

**Fig. 2 | Temperature and current-amplitude dependences of second-harmonic magnetoresistance in Sm1111 with $\nabla n_\mathrm{H}$. a**, **b** Magnetic-field dependence of the first- and second-harmonic resistance ($R_{xx}^{\omega}$ and $R_{xx}^{2\omega}$, respectively) of sample #1 for various temperatures across the superconducting transition ($T_c^{\mathrm{onset}} = $ 41 K). As shown in the inset schematic figure, the magnetic field and electric current are applied mutually orthogonally and both perpendicular to the hydrogen concentration gradient. The amplitude of input ac current was 0.5 mA. For clarity, curves in (**b**) are vertically offset by 10 or 20 mΩ. **c**, Magnetic-field dependence of $R_{xx}^{2\omega}$ at $T = 33$ K under various current amplitudes ranging from 0.1 to 5 mA. The magnetic-field and current geometries are identical to those in the inset of Fig. 2a. **d**, Current amplitude dependence of $R_{xx}^{2\omega}$ under $B = 1.95$ T at $T = 33$ K. Dashed line is a guide to the eye indicating the linear relationship between $R_{xx}^{2\omega}$ and current amplitude at low-current regime. The inset displays an enlarged view of the low-current regime.

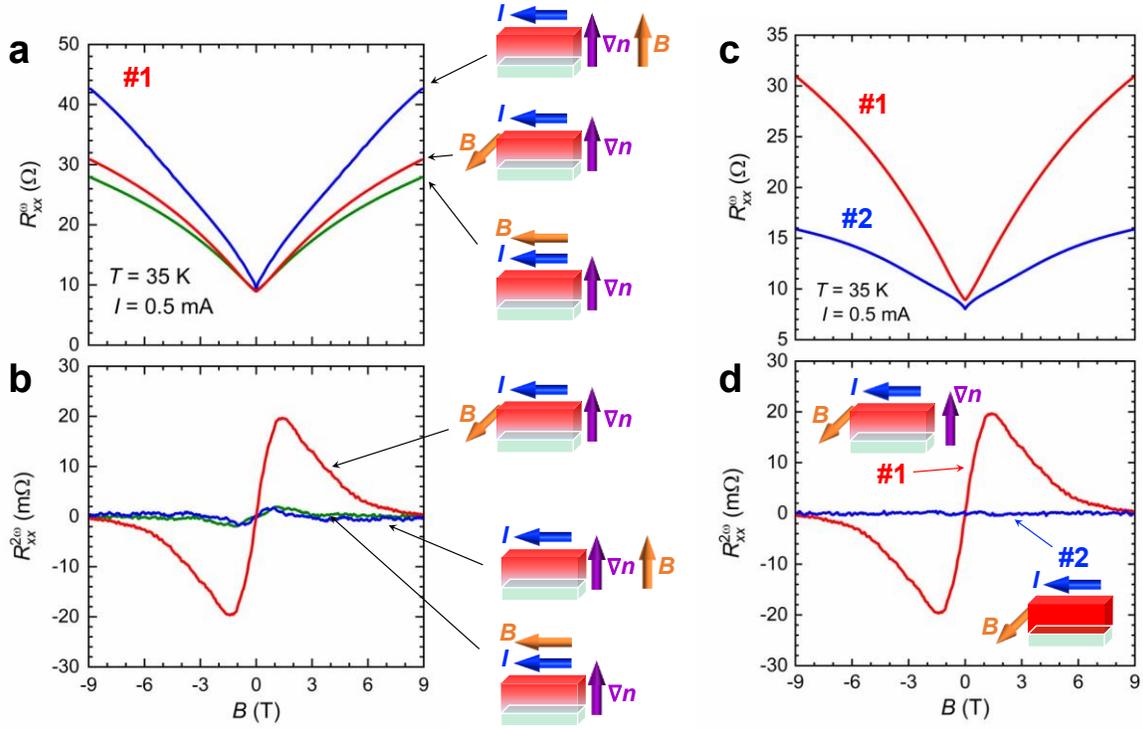

**Fig. 3 | Geometry and concentration-gradient dependences of nonreciprocal charge transport. a, b,** magnetic-field dependence of $R_{xx}^{\omega}$ and $R_{xx}^{2\omega}$ of sample #1 at $T$ = 33 K with $I_{ac}$ = 0.5 mA, measured under various magnetic-field and current configurations: (i) $\boldsymbol{I} \perp \boldsymbol{B} \perp \boldsymbol{\nabla} n_{H}$, (ii) $\boldsymbol{I} \perp \boldsymbol{B} \parallel \boldsymbol{\nabla} n_{H}$, (iii) $\boldsymbol{I} \parallel \boldsymbol{B} \perp \boldsymbol{\nabla} n_{H}$. **c, d,** Comparison of the magnetic-field dependences of the $R_{xx}^{\omega}$ and $R_{xx}^{2\omega}$, between sample #1 (with $\boldsymbol{\nabla} n_{H}$) and sample #2 (without $\boldsymbol{\nabla} n_{H}$). The field and current were arranged as configuration (i) $\boldsymbol{I} \perp \boldsymbol{B} \perp \boldsymbol{\nabla} n_{H}$ with $I_{ac}$ = 0.5 mA at $T$ = 35 K.

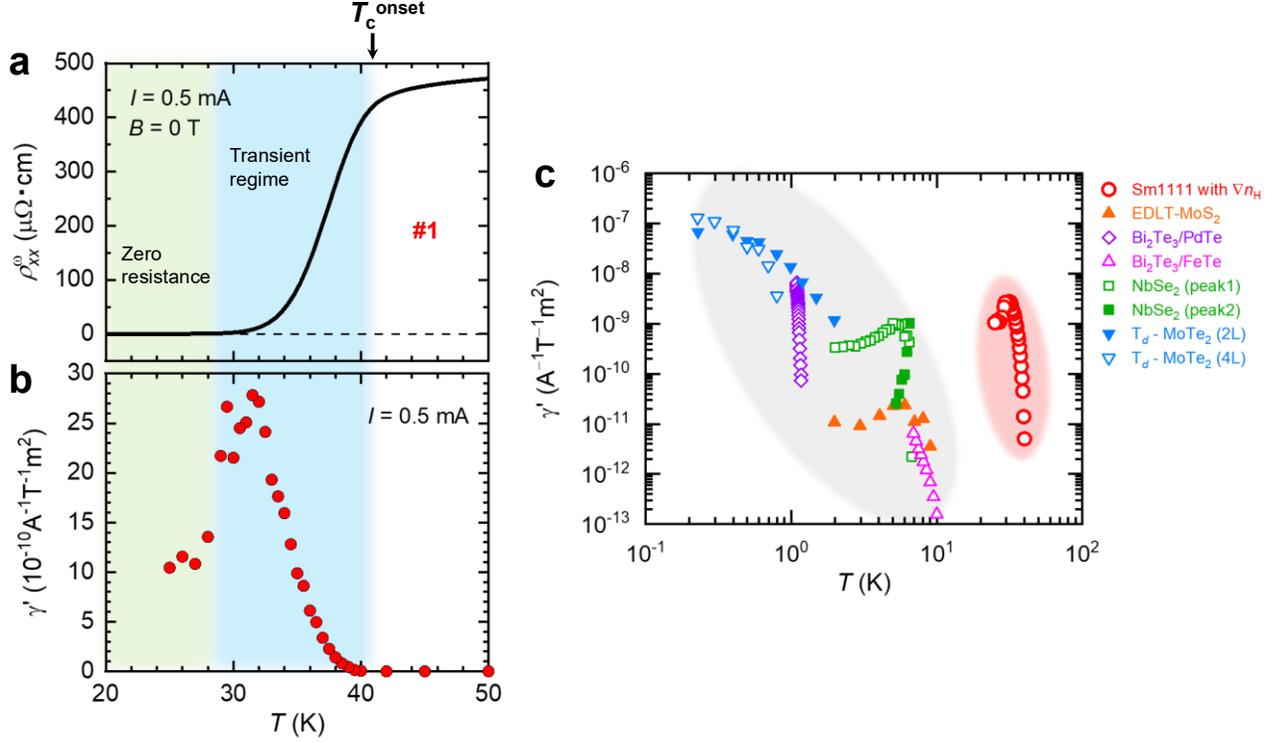

**Fig. 4 | Temperature evolution of nonreciprocal coefficient and comparison across systems. a,b,** Temperature dependence of first-harmonic resistivity $\rho_{xx}^{\omega}$ and nonreciprocal coefficient $\gamma$ of sample #1 under ac current amplitude $I_{ac}$ = 0.5 mA. We adopt $\gamma = \frac{2R_{xx}^{2\omega}}{R_{xx}^{\omega}BI_0}$, evaluated from the slope of the $R_{xx}^{2\omega}/R_{xx}^{\omega}$-$B$ profile near 0 T, deviated from Figs. 2**a**, **b**. (see Supplementary Note 2). The green shading roughly denotes the full superconducting state with zero resistance, as a guide to the eye. The blue shading indicates the transient regime between $T_c^{onset}$ (onset of the resistivity drop) and the zero-resistance temperature. **c**, Size-independent nonreciprocal coefficient $\gamma'$ as a function of temperature for superconductors exhibiting vortex-motion-driven nonreciprocal charge transport [33, 39, 36, 38, 57]. The data of Sm1111 with $\nabla n_H$ studied here are marked by open red circles. To estimate the value of $\gamma'$ for interfacial two-dimensional superconductors EDLT-MoS$_2$ [33], Bi$_2$Te$_3$/PdTe [39], and Bi$_2$Te$_3$/FeTe [36], we adopt an effective superconducting thickness of 1 nm, following Ref. [45].

# Supplementary Information

**Supplementary Note 1 | Nonreciprocal charge transport measurement**

Phenomenologically, electrical resistance can be expanded up to second order in field and current as

$$R(I, B) = R_0 \left(1 + \beta B^2 + \gamma BI\right) \tag{S1}$$

where $R_0$, $I$, and $B$ are the zero-field resistance, the applied current, and the magnetic field, respectively [1, 2]. The second term with coefficient $\beta$ represents the ordinary magnetoresistance. The third term depends on the direction of $I$ and $B$ and produces a nonreciprocal component, as seen from $\Delta R = R(B, I) - R(B, -I) \propto \gamma BI$, which is odd between $I$ and $-I$. In centrosymmetric systems, the third term is forbidden by symmetry, implying $\gamma = 0$ if the crystal retains inversion symmetry. Observing a finite nonreciprocal response therefore provides an electrical probe of inversion symmetry breaking.

For a polar system, symmetry refines Eq. (S1) to

$$R(I, B) = R_0 \left[1 + \beta B^2 + \gamma (\boldsymbol{B} \times \hat{\boldsymbol{z}}) \cdot \boldsymbol{I}\right] \tag{S2}$$

where $\hat{\boldsymbol{z}}$ is the unit vector along the polar axis. The current-dependent resistance generates a nonlinear voltage drop that can be detected by second-harmonic lock-in techniques under an ac current $I(t) = I_{ac} \sin \omega t$ [3]. The voltage originating from the third term yields

$$V^{2\omega}(t) = \gamma R_0 B I_{ac}^2 \cos \theta \sin^2 \omega t$$

$$= \frac{1}{2} \gamma R_0 B I_{ac}^2 \cos \theta \left\{1 + \sin\left(2\omega t - \frac{\pi}{2}\right)\right\} \tag{S3}$$

where $\theta$ is the angle between $\boldsymbol{B}$ and $\boldsymbol{I}$. Consequently, in the low-field limit, the first- and second-harmonic resistances become

$$R^\omega \equiv \frac{V^\omega}{I_{ac}} \approx R_0, \quad R^{2\omega} \equiv \frac{V^{2\omega}}{I_{ac}} = \frac{1}{2} \gamma R_0 B I_{ac} \cos \theta.$$

In our experiments we detect a lock-in $y$-component of the second-harmonic voltage with a $\pi/2$ phase shift, and confirmed that the $x$-component was almost zero.

**Supplementary Note 2 | Evaluation of the nonreciprocal coefficient $\gamma$**

For the orthogonal geometry with $\hat{z} \perp \boldsymbol{B} \perp \boldsymbol{I}$, Eq. (S2) gives

$$\frac{R^{2\omega}}{R^{\omega}} = \frac{1}{2}\gamma I_{ac}B.$$

Thus, $\gamma$ is obtained from the slope of the low-field linear region for $R^{2\omega}/R^{\omega}$-$B$ curve. Figure S1 shows representative $R^{2\omega}/R^{\omega}$-$B$ curves at several temperatures across the superconducting transition temperature $T_c$. We evaluate $\gamma$ by least-squares linear fitting around $\boldsymbol{B} = 0$.

**Supplementary Note 3 | Symmetrisation and antisymmetrisation of raw data**

To separate the even and odd components with respect to applied magnetic fields, we symmetrise/antisymmetrise the raw magnetoresistance $R_{exp}(B)$ measured at positive and negative fields:

$$R_{sym} = \frac{R_{exp}(B) + R_{exp}(-B)}{2}, \quad R_{asym} = \frac{R_{exp}(B) - R_{exp}(-B)}{2}.$$

$R_{sym}(B)$ is even and $R_{asym}(B)$ is odd in $B$. We apply the same procedure to both $R^{\omega}$ and $R^{2\omega}$ throughout the analysis. Typical raw and symmetrised/antisymmetrised datasets are shown in Fig. S2. It is noted that the small residual even components for $R^{2\omega}$ likely arise from slight contact-resistance asymmetries.

**Supplementary Note 4 | Basic transport properties of prepared samples**

Figures S3a, b display the temperature dependence of the electrical resistivity for hydrogen-doped Sm1111 epitaxial thin films: sample #1 (with a concentration gradient) and sample #2 (uniform hydrogenation). In the main manuscript, $T_c^{onset}$ is defined as the intersection of two linear fits, one to the normal state and the other to the transient region.

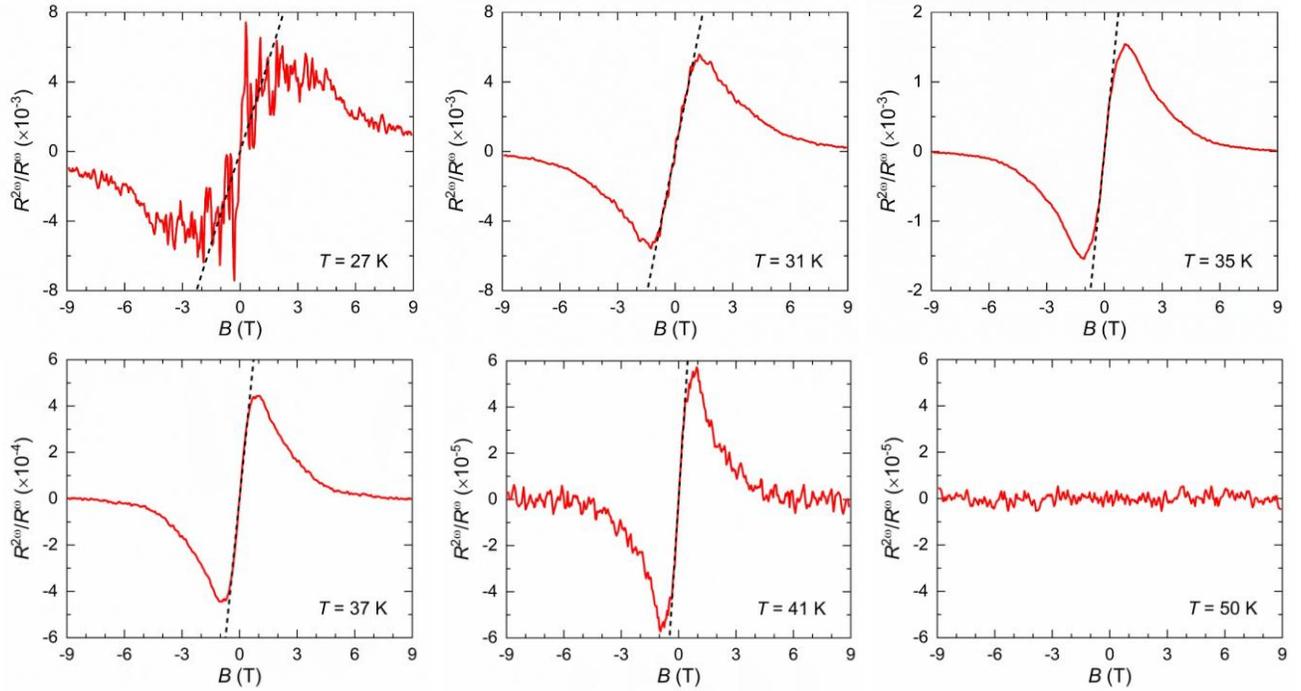

**Supplementary Fig. S1 | Evaluation of nonreciprocal coefficient $\gamma$.** Magnetic field dependence of $R_{xx}^{2\omega}/R_{xx}^{\omega}$ measured under $I_{ac}$ = 0.5 mA for temperatures across $T_c$. The dashed lines indicate linear fits to the low-field regime. The nonreciprocal coefficient $\gamma$ is obtained from the slope of the $R_{xx}^{2\omega}/R_{xx}^{\omega}$-$B$ curve around 0 T. At 50 K above the transition temperature ($T_c^{onset}$ = 41 K), $R_{xx}^{2\omega}$ is effectively zero (below the detection limit); consequently, $R_{xx}^{2\omega}/R_{xx}^{\omega}$ shows no discernible signal.

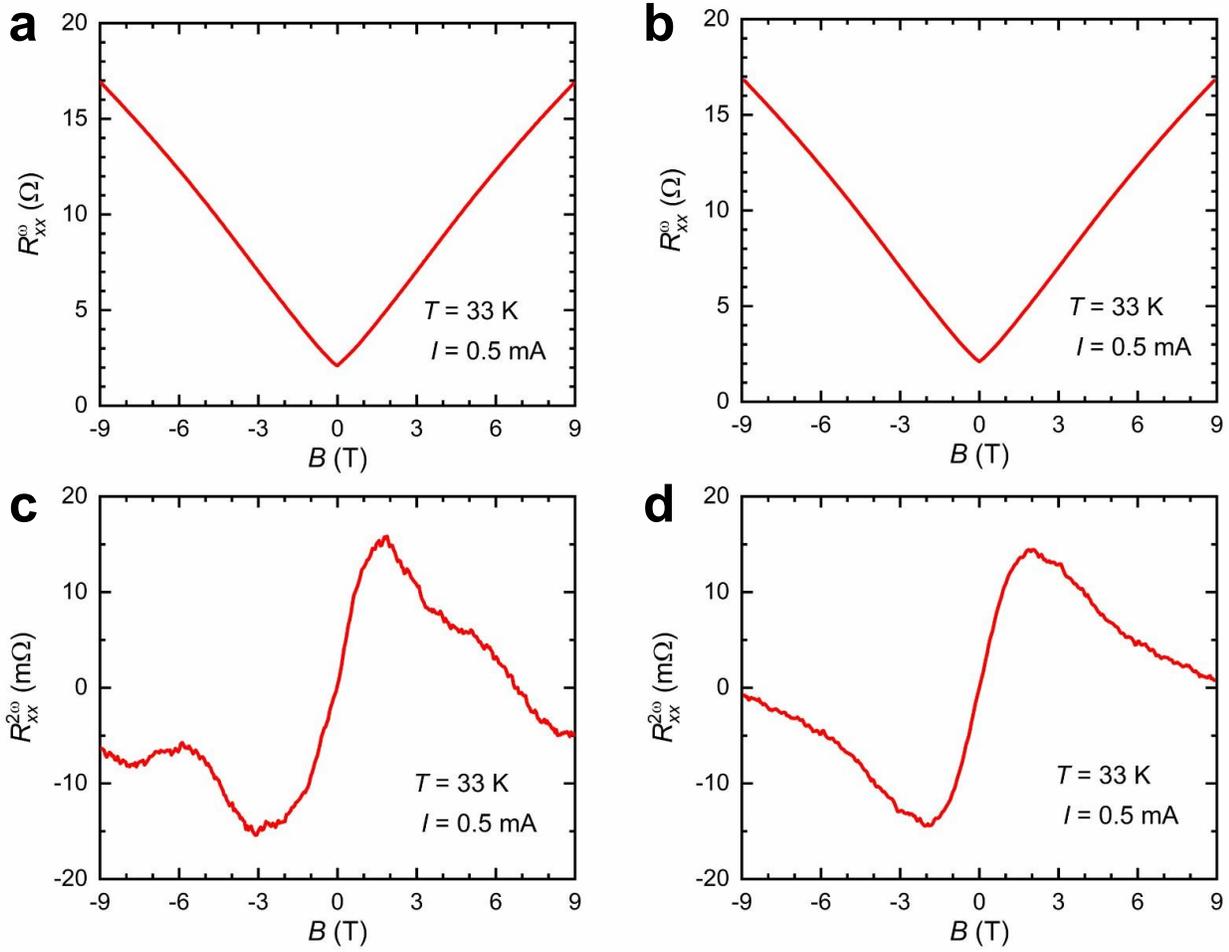

**Supplementary Fig. S2 | Symmetrisation and antisymmetrisation of first- and second-harmonic magnetoresistance. a**, **b** Raw (a) and symmetrised (b) first-harmonic magnetoresistance $R_{xx}^{\omega}$ under $I_{ac}$ = 0.5 mA at 33 K. **c**, **d** Raw (c) and antisymmetrised (d) second-harmonic magnetoresistance $R_{xx}^{2\omega}$ under $I_{ac}$ = 0.5 mA at 33 K.

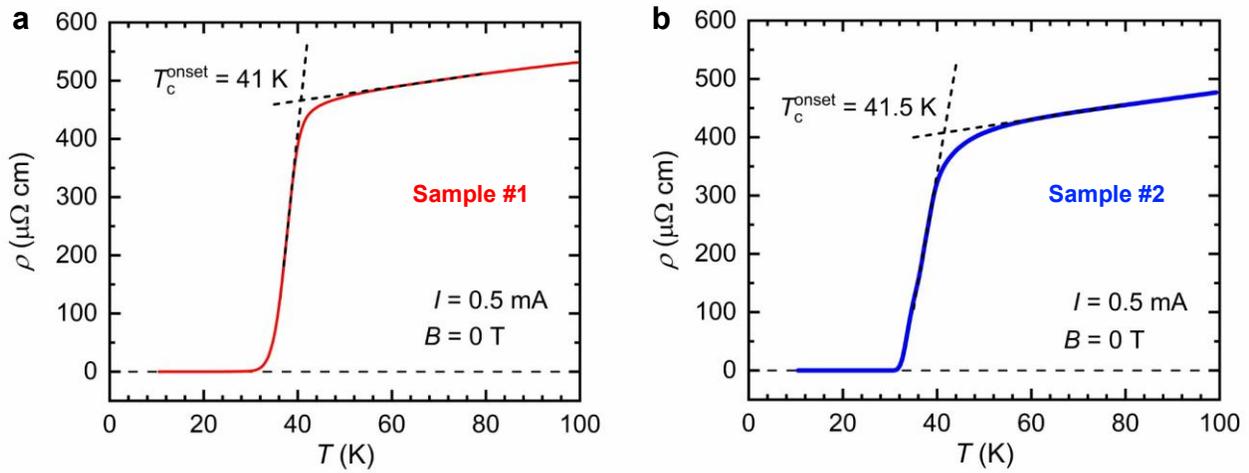

**Supplementary Fig. S3 | Basic transport properties of the prepared samples.** Temperature dependence of electrical resistivity $\rho$ of (a) sample #1 (with a hydrogen concentration gradient) and (b) sample #2 (without a gradient) measured under dc current $I_{dc}$ = 0.5 mA. The onset of superconducting transition $T_c^{onset}$ is evaluated from the intersection of two linear fits to the $\rho$-$T$ curve (normal-state and transient-regime segments).

**Supplementary References**